\documentclass[apj,twocolumn]{openjournal}
\usepackage{tikz}
\usetikzlibrary{arrows.meta,decorations.pathmorphing,calc,positioning}
\usepackage{newtxtext,newtxmath,enumitem,multirow,ctable}
\usepackage[breaklinks,colorlinks,citecolor=blue,urlcolor=blue]{hyperref}
\usepackage{comment}

\newcommand{\rp}{r_{\rm p}}
\newcommand{\rtidal}{r_{\rm t}}

\newcommand{\MBH}{M_{\bullet}}

\newcommand{\orcidauthor}[3]{\author{\href{http://orcid.org/#1}{#2$^{#3}$}}}

\shorttitle{Mass Transfer in Tidally Heated Stars}
\shortauthors{Yao \& Quataert}

\begin{document}

\title{Mass Transfer in Tidally Heated Stars Orbiting Massive Black Holes\\ and Implications for Repeating Nuclear Transients\vspace{-4em}}

\orcidauthor{0000-0003-3024-7218}{Philippe Z. Yao}{1*}
\orcidauthor{0000-0001-9185-5044}{Eliot Quataert}{1}
\affiliation{$^{1}$Department of Astrophysical Sciences, Princeton University, Princeton, NJ 08544, USA}
\thanks{$^*$E-mail: \href{mailto:philippe.yao@princeton.edu}{philippe.yao@princeton.edu}}

\begin{abstract}
The structure of stars orbiting close to supermassive black holes (SMBHs) can be dramatically modified by tidal heating, which can in principle dissipate an energy much larger than the stellar binding energy.  We use analytic models and MESA  to explore the coupled dynamics of tidal heating,  stellar structural evolution, orbital decay due to gravitational waves and tides, and mass transfer.   In contrast to more equal mass stellar binaries, the stable mass transfer rate for stars orbiting SMBHs is typically set by the tidal heating timescale (the timescale for tides to increase the stellar radius), not by the gravitational wave orbital decay timescale.  The resulting stable mass transfer rate is sensitive to the tidal heating model but is plausibly $\sim 10^{-5}-10^{-3} M_\odot {\, \rm yr^{-1}}$ (and perhaps larger), sufficient to produce low-luminosity active galactic nuclei in many galaxies.  The stability of mass transfer is sensitive to {\em where} in the stellar interior the tidal energy is dissipated. MESA models confirm the expected result that mass transfer is unstable (stable) if tidal heating increases (decreases) the fraction of the star that is convective.   More detailed conclusions about the stability of mass-transfer will require self-consistently calculating how the tidal heating of stars changes in response to internal structural changes produced by the tidal heating itself. Stars with tidal heating-induced mass transfer can produce a large population of low-luminosity active galactic nuclei; they may also be the progenitors of some partial tidal disruption candidates (e.g., ASASSN-14ko) as well as short-period quasi-periodic eruptions (e.g.,  eRO-QPE2 and GSN 069).  However, many repeating nuclear transients produced by tidal heating-induced mass loss are likely fainter than those detected thus far, and remain to be discovered.
\end{abstract}

\keywords{binaries: general --- quasars: supermassive black holes --- stars: evolution --- stars: mass-loss --- transients: tidal disruption events}

\maketitle

\section{Introduction}
\label{sec:intro}
The fate of stars orbiting close to supermassive black holes depends on the complex interplay of gravitational interactions and hydrodynamic processes such as tides, stellar collisions, and star-accretion disk collisions. For example, stars in a nuclear cluster can be scattered onto orbits where the pericenter passage intercepts the tidal radius, resulting in tidal disruption events (TDEs) \citep[e.g.][]{Hills1975, Rees1988}. While TDEs have been systematically studied with over 100 candidates detected so far \citep[e.g.][]{Gezari2021, Yao2023}, their less extreme counterparts have become a recent focus of attention with the discovery of nearly a dozen repeating nuclear transients (RNTs) in the past few years by all-sky optical and X-ray surveys.

Multiple types of RNTs have been observed, including quasi-periodic eruptions (QPEs) and partial-tidal disruption candidates (pTDEs). QPEs are recurring soft X-ray flares with intervals ranging from hours to days, exhibiting diverse timing patterns and lacking bright optical or UV counterparts \citep[e.g.][]{Miniutti2019, Arcodia2021, Chakraborty2021, Chakraborty2024, Chakraborty2025, Arcodia2022, Miniutti2023, Webbe2023, Arcodia2024}. In contrast, pTDEs such as ASASSN-14ko \citep{Payne2021}, eRASSt J045650.3-203750 \citep{Liu2023eRASSt}, and AT2022dbl (\citealt{Hinkle2024}; Makrygianni et al. submitted) produce flares in the optical, UV, and/or X-ray with a significantly longer mean recurrence time. For ASASSN-14ko, with a mean period of $P_0 = 114.2\pm 0.4$ days and flare duration of $\sim 10$ days, a period derivative is also observed at $\dot P \approx -0.0026 \approx -(7{\rm h})/P_0$ \citep{Payne2022}, significantly more rapid orbital evolution than can be produced by gravitational wave inspiral of a star \citep{Payne2021}. 

Observations of RNTs are suggestive of their origin in a binary system consisting of a star orbiting a supermassive black hole (SMBH) (see, e.g., \citealt{Xian2021, Sukova2021,Krolik2022, LinialMetzger2023,Linial2024b,Lu2023, Franchini2023,Liu2023,Liu2025, LQ2024, Bandopadhyay2024,Yao2024}). In such models, the observed differences in the properties of RNTs are primarily due to the stellar mass, black hole mass, and orbital parameters such as inclination, eccentricity, and orbital period. For QPEs, the alternating long-short time between flares in GSN 069 and eRO-QPE2 \citep{Miniutti2019,Miniutti2023,Arcodia2021} suggests a star on a mildly eccentric orbit ($e\lesssim 1$) with a period of hours to days. Host galaxy studies also indicate that the black hole masses are $\sim 10^5 - 10^6 M_{\odot}$ \citep[e.g.][]{Wevers2022}. In contrast, pTDEs may arise from stars on highly eccentric orbits ($1-e\ll1$) with periods of months to years \citep{Payne2021, Cufari2022, Liu2023,Bandopadhyay2024,Liu2025}, likely powering less frequent yet more energetic repeating transients.

Even if the pericenter distance of the star is outside its tidal radius, the orbiting star can be strongly affected by the tidal force of the SMBH via dissipation of tidal energy. Previous work has shown that interior to a pericenter distance of $\sim 4-5 \times$ the tidal radius ($\rtidal = R_\star (\MBH/M_\star)^{1/3}$), tidal heating can lead to runaway expansion of the stellar radius, initiating mass transfer onto the central black hole \citep{Li2013, Lu2021, Linial2024Period,LQ2024}.   Indeed, it is likely inevitable that tidal heating changes the structure of most stars approaching a SMBH secularly, e.g., by gravitational wave orbital decay or by small scattering-induced changes in pericenter distance for stars on highly eccentric orbits -- the empty loss cone regime of stellar dynamics in galactic nuclei. Additionally, it is plausible that the presence of an ambient accretion disk, either from another tidally disrupted star or as a result of tidal heating-induced mass transfer, can further inflate the stellar envelope and remove mass from the star via repeated collisions with the disk.  We refer the reader to \citet{Yao2024} for a detailed discussion of star-disk interactions, which we do not model in this paper.

In this paper, we build on previous work on tidally heated stars near SMBHs to study the fate of these stars both analytically and in 1D \texttt{MESA} simulations. We focus for concreteness on Sun-like stars on either mildly or highly eccentric orbits around SMBHs. Here, “Sun-like stars” refers to main-sequence stars with an outer convective envelope surrounding an inner radiative zone.  As in more equal mass binaries, there are likely regimes of both stable and unstable mass transfer \citep[e.g.][]{Paczynski1971, Hjellming1987, Ge2015}.   An important difference between the high mass ratio case relevant to stars around SMBHs and the more equal mass case of stellar binaries is that the high mass ratio makes it more likely that mass transfer is unstable \citep{linialSari2023,Lu2023}.   This is because it is difficult for the angular momentum lost from the star via mass transfer to return to the orbit, for three separate reasons: (1) tidal torques between the star and the disk are inefficient at high mass ratio (analogous to the inability of Earth-mass objects to open gaps in protostellar disks), (2) mass transfer via L2 is efficient at high mass ratio and extracts angular momentum from the orbit \citep{2017MNRAS.469.2441L}, (3) the proximity of the tidal radius to the last stable circular orbit implies that a significant amount of angular momentum can be accreted by the black hole.   Taken together these results suggest that the star's orbit may not evolve significantly in response to mass loss.  In this case, the stability of mass transfer is largely set by the star's response to mass-loss:  if $R_\star \propto M_\star^\zeta$ with $\zeta \lesssim 1/3$, the mass transfer will likely be unstable \citep{2017MNRAS.469.2441L}.   This suggests that mass-transfer by fully convective stars is unstable, while that by radiative stars is stable \citep{Lu2023}.  The two regimes of mass transfer for stars orbiting SMBHs will have very different observational outcomes. For example, stable mass transfer may power long-lived low-luminosity active galactic nuclei \citep{Hameury1994, MacLeod2013}. On the other hand, unstable mass transfer would produce a runaway process that leads to super-Eddington accretion, appearing as a transient flare akin in some ways to TDEs \citep{LQ2024}.

The primary goal of this paper is to identify qualitatively new features of mass-transfer in tidally heated stars and assess their possible observational implications.  Solving the full coupled problem of orbital and stellar evolution for stars near a BH's tidal radius is formidable, so it is worth stressing up front a few of the assumptions and limitations of our analysis.  These simplifying assumptions are made largely because of uncertainties in the key tidal heating physics.   The first is that we will use models of tidal heating that depend only on the global properties of the star (mass, radius) and its pre-tidal-heating stellar structure.   This is in general not a good assumption because tidal heating changes the stellar structure significantly (as we shall show), which in turn changes the tidal excitation of waves and the resulting tidal heating rate.   Secondly, we will not attempt to calculate in detail where in the star the tidal energy is dissipated, despite this being important for determining how the star responds to tidal heating.   Instead, we will use phenomenological models for how tidal energy is spatially distributed.    And, finally, our analysis will include both mildly eccentric and highly eccentric orbits.   The reader should be warned up front that our theoretical tools (analytics and MESA) are better suited to lower eccentricity orbits, but we believe they also provide qualitative insight into the role of tidal heating for stars on highly eccentric orbits. We return to prospects for relaxing these assumptions in \S \ref{sec:summary}.

The remainder of this paper is organized as follows. We outline our analytic model and show the resulting orbital and stellar evolution for a Sun-like star in \S \ref{sec:analytic}. We then carry out similar calculations conducted with more realistic stellar models with \texttt{MESA} in \S \ref{sec:MESA}. Our simulations cover a wide range of orbital parameters and tidal heating models, and are evolved until unstable mass transfer commences. We explore the implications of the stable and unstable mass transfer phases seen in our calculations for recently discovered RNTs and provide a discussion and summary of our models and findings in \S \ref{sec:summary}.

\section{Stable Mass Transfer with Tidal Heating}
\label{sec:analytic}

Consider a star of mass $M_{\star}$ in orbit around a supermassive black hole of mass $\MBH \gg M_{\star}$.   The orbital period is $P$, the star's eccentricity is $e$ and the pericenter distance $\rp$ is comparable to the tidal radius $\rtidal \simeq R_{\star} (\MBH/M_{\star})^{1/3}$. Gravitational wave radiation causes orbital decay on a timescale $\tau_{\rm GW}$, eventually leading to mass transfer when $\rp = 2r_{\rm t}$. Considering first a case without tidal heating, the stable mass transfer rate would be set by angular momentum loss driven by gravitational wave radiation. From \citet{Peters1964}, for highly eccentric ($1-e \ll 1$) and near-circular ($e \ll 1$) orbits, this timescale can be approximated as:

\begin{equation}
\begin{aligned}
\tau_{\rm GW,para} &\sim \frac{\rp}{|\dot \rp|} = \frac{24\sqrt{2}}{59} \frac{c^5}{G^3}\frac{\rp^{5/2}a^{3/2}}{\MBH^2 M_{\star}}\\
\tau_{\rm GW,circ} &\sim \frac{a}{|\dot a|} = \frac{5}{64}\frac{c^5}{G^3}\frac{a^4}{\MBH^2 M_{\star}}
\label{eq:tgrav}
\end{aligned}
\end{equation}

The resulting stable mass transfer solution has a mass transfer rate from the star to the BH of  $\dot M \sim M_{\star}/\tau_{\rm GW}$

\begin{equation}
\begin{aligned}
    \dot M_{\rm GW,para} &\simeq  4.6\times10^{-10} M_{\odot}  {\rm yr}^{-1} \left(\frac{R_{\star}}{R_{\odot}}\right)^{-5/2}\\
    &\ \ \ \left(\frac{M_{\star}}{M_{\odot}}\right)^{17/6} \left(\frac{\MBH}{10^6 M_{\odot}}\right)^{2/3}\left(\frac{P}{\rm 100 days}\right)^{-1}\\
\dot M_{\rm GW,circ} &\simeq  10^{-6} M_{\odot} {\rm yr}^{-1} \left(\frac{R_{\star}}{R_{\odot}}\right)^{-4} \left(\frac{M_{\star}}{M_{\odot}}\right)^{10/3} \left(\frac{\MBH}{10^6 M_{\odot}}\right)^{2/3}
\label{eq:MdotGW}
\end{aligned}
\end{equation}

The above calculation neglects the fact that tides raised in the star by the BH can deposit sufficient energy to dramatically change the structure of the star before mass transfer is initiated at $\rp = 2\rtidal$. Heat deposited inside the star increases its radius when the integrated tidal energy deposited is a fraction of the stellar binding energy (as we shall see, the heating timescale near $\rtidal$ is less than the thermal timescale of the low mass stars we focus on in this paper so the tidal energy cannot be efficiently radiated away). Because the orbital energy of a star near $\rtidal$  is a factor of 
\begin{equation}
\frac{E_{\rm orb}}{E_\star}\sim \frac{G\MBH M_\star/(2r_{\rm t})}{GM_\star^2/R_\star}\sim(\MBH/M_\star)^{2/3}
\label{eq:EorbEstar}
\end{equation}
 larger than its binding energy, tidal energy dissipation of just a small fraction of the orbital energy can unbind the star.  \citet{LQ2024} showed that standard estimates of tidal heating predict that the tidal energy dissipated on the gravitational wave inspiral time is indeed large enough to unbind the star (see also \citealt{Li2013, Lu2021} for related calculations).   These results imply that tidal heating inflates the radius of the star, leading to mass transfer earlier than would be expected absent tidal heating.  For now, we assume that the resulting mass transfer is stable (although this is by no means guaranteed), and quantify the properties of this unusual regime of stable mass transfer.   In the presence of strong tidal heating,  the stable mass transfer solution is no longer set by $\tau_{\rm GW}$, but is instead set by the tidal heating timescale $\tau_{\rm heat} \sim E_{\star}/\dot E_{\rm tides}$ when $\tau_{\rm heat} < \tau_{\rm GW}$, where $E_{\star}$ is of order the stellar binding energy and $\dot E_{\rm tides}$ is the tidal heating rate. 

\subsection{Effects of Tidal Heating} \label{sec:tides_noGW}

In what follows, we consider stars on both roughly circular ($e \ll 1$) and highly eccentric ($1-e \ll 1$) orbits (the results for tidal heating are cleanest in these two regimes). For stars on highly eccentric orbits, the black hole's tidal potential excites significant internal stellar oscillation modes at each pericenter passage. The amount of heating in this regime can be estimated via the linear perturbation theory of \citet{Press1977}. We assume that non-linear coupling between modes excites an abundance of daughter modes after pericenter passage, which quickly dissipates the tidally excited modes \citep{Kumar1996, Linial2024Period}. When this is the case and the damping time of the tidally excited oscillations is much shorter than the orbital period, the time-averaged tidal heating rate can be approximated as 
\begin{equation}
\dot E_{\rm tides,p} \simeq \frac{GM_{\star}^2/R_{\star}}{P} f(\chi)
\label{eq:Edot_para}
\end{equation}
where $\chi \equiv \rp/\rtidal$. For a Sun-like star on a highly eccentric orbit near mass transfer ($\chi \sim 2$), we take $f(\chi) \simeq 0.33 f_{\rm p}\chi^{-\alpha}$ with $\alpha \simeq 15$ based on the calculations in  \citet{Lai1997} which generalize those of \citet{Press1977} and \citet{Lee1986} to a rotating star.   In this expression for $f(\chi)$, $f_{\rm p}$ is a dimensionless number intended to encapsulate the uncertainty in the tidal heating theory for highly eccentric orbits.   The expression here for the energy transfer assumes that the star has been spun up to the point where its rotation rate no longer evolves significantly due to tides.  This is a reasonable assumption because the tidal spin-up time is shorter than other key timescales in the problem (e.g., the GW inspiral time, the tidal heating time, etc).   A useful numerical calibration of the linear theory results leading to equation \ref{eq:Edot_para} is provided by the simulations of \citet{Cufari2023}, who simulated the tidal energy deposited in an $n = 3/2$ polytrope and found that the energy deposited per orbit is $\Delta E \simeq 0.02 GM_\star^2/R_\star$ for $r_p \simeq 2 r_t$.   This is only a factor of few higher than the linear tidal calculations of \citet{Lee1986}. \citet{Cufari2023}'s tidal heating rate is significantly larger than our default $f(\chi)$ from \citet{Lai1997} in part because solar-type stars are better approximated by $n = 3$ polytropes and in part because the tidal energy deposition decreases when the star is rapidly rotating, especially when the stellar spin becomes comparable to the orbital angular frequency near pericenter for prograde orbits \citep{Lai1997,Kumar1998}.

When the non-linear dissipation time of the tidally excited modes is longer than the orbital period, the instantaneous tidal energy input rate and heating rate are not given by eq \ref{eq:Edot_para}. Instead, the tidally excited modes can be chaotically driven to even larger amplitudes \citep[e.g.][]{Mardling1995a,Mardling1995b,Wu2018,Vick2018}. In this regime, the average tidal heating rate per orbit is still given by equation \ref{eq:Edot_para}.

For a solar-type star with a radiative core and a convective envelope, dynamical tidal dissipation through internal gravity waves can also produce significant heating inside the star \citep{Goodman1998}. In the case of a low-eccentricity and short-period orbit of a solar-mass star around a massive BH, the tidal heating rate can be estimated via equation A7 in \citet{LQ2024}:
\begin{equation}
\dot{E}_{\text{tides,c}} \simeq 4 \times 10^{38} \, \text{erg s}^{-1} f_{\rm c} e^2 \left( \frac{a}{2r_{\text{t}}} \right)^{-23/2}, 
\label{eq:Edot_circ}
\end{equation}
where $f_{\rm c}$ is a dimensionless number intended to capture the uncertainty in the tidal heating theory for circular orbits.  In particular, inertial wave excitation in convection zones can potentially lead to $f_c \gg 1$ \citep{Barker2020}, so we will consider that possibility as well in some of our calculations (\citealt{LQ2024} estimated $f_c \sim 100$ for inertial wave excitation in solar-type stars).   Note that equation \ref{eq:Edot_circ} should in general include a dependence on the stellar mass and the internal structure of the star (e.g., its evolutionary state or its tidally-modified structure).   We neglect this dependence for this initial exploration, but return to this point in \S \ref{sec:summary}.

For the analytic estimates in this section, we neglect radiative losses, motivated by the fact that the tidal heating  timescale near $\rtidal$ is typically shorter than the thermal time of low mass stars (this assumption will be relaxed in our MESA models in \S \ref{sec:MESA}).  In this case, we can calculate the evolution of the star's properties via
\begin{equation}
\frac{d E_{\rm tot}}{dt} = -\phi_E \frac{d}{dt} \frac{G M_{\star}^2}{R_{\star}}= \dot E_{\rm tides}
\label{eq:tideode}
\end{equation}
where $E_{\rm tot}$ is the total energy of the star, and $\phi_E$ is a dimensionless structural coefficient that links the star's total energy to the simple scale $E_{\star}\equiv GM_{\star}^2/R_{\star}$, which is of order the binding energy of the star. For a Sun-like star, we obtained $\phi_E \simeq 0.82$ from a \texttt{MESA} model.

When the total tidal heating becomes a fraction of the stellar binding energy, the stellar structure will gradually be modified, reflected primarily by an expansion in radius. Prior to the onset of mass transfer, the radius of the star will increase in response to significant tidal heating, implying that mass loss then becomes inevitable. Once the star begins transferring mass to the BH, equation \ref{eq:tideode} becomes
\begin{equation}
\phi_E \left(\frac{d \ln R_{\star}}{dt} - 2 \frac{d \ln M_{\star}}{dt}\right) = \frac{\dot E_{\rm tides}}{E_{\star}}.
\label{eq:tideode2}
\end{equation}
We now assume, following \citet{Lu2023} and \citet{linialSari2023}, that once mass transfer commences, the angular momentum and energy lost from the star via mass loss are permanently lost and cannot be returned to the orbit (as discussed in the introduction, the authors of these studies argue that this is appropriate for the high mass ratio limit of a star around a massive BH). This implies that the orbit does not change in response to mass loss.  Since $r_p$ is fixed, maintaining $\rp \simeq 2\rtidal$ as $M_{\star}$ and $R_{\star}$ change requires $R_{\star} \propto M_{\star}^{1/3}$.    This is the stable mass transfer solution, specialized to the case of constant $\rp$; stable mass transfer further assumes that the stellar radius does not intrinsically expand more rapidly than $R_{\star} \propto M_{\star}^{1/3}$ in response to mass loss (we study this in MESA in \S \ref{sec:MESA}). If we substitute $R_{\star} \propto M_{\star}^{1/3}$ into equation \ref{eq:tideode2} and use $r_p = 2 r_t$, we can solve an ODE to obtain the mass transfer rates for the two types of orbits
\begin{equation}
    \begin{aligned}
    \dot M_{\star,\rm para}(t) &= -C_0 M_{\star,\rm MT}\, {\rm exp}(-C_0 t)\\
    \dot M_{\star,\rm circ}(t) &= -C_1\left(M_{\star,\rm MT}^{5/3} - \frac{5}{3}C_1 t\right)^{-2/5},\\
    \end{aligned}
    \label{eq:MdotTides}
\end{equation}
where
\begin{equation}
    \begin{aligned}
    C_0 = \frac{1}{5}\frac{f_{\rm p}\chi_{\rm MT}^{-15}}{P\phi_{\rm E}};\ \ C_1 = \frac{3}{5}\frac{L_0 f_{\rm c}}{\phi_{\rm E} G}\frac{R_{\star,\rm MT}}{M_{\star,\rm MT}^{1/3}}e^2(1-e)^{23/2},
    \end{aligned}
    \label{eq:MdotTidesConstants}
\end{equation}
$L_0 = 4\times10^{38} \rm \, erg \, s^{-1}$ (Eq. \ref{eq:Edot_circ}), MT refers to stellar or orbital properties at the onset of mass transfer, and $t$ measures the time since the star begins to lose mass.  Evaluated at this time, the mass transfer rate is
\begin{equation}
    \begin{aligned}
    \dot M_{\star,\rm para}(t_{\rm MT}) &= 7.5\times10^{-5}  M_{\odot} {\rm yr}^{-1} \frac{M_{\star,\rm MT}}{M_{\odot}} \left(\frac{P_{\rm MT}}{100\rm \, days}\right)^{-1}f_{\rm p}\\
    \dot M_{\star,\rm circ}(t_{\rm MT}) &= 2.4\times10^{-3} M_{\odot} {\rm yr}^{-1} \frac{R_{\star,\rm MT}}{R_{\odot}} \left(\frac{M_{\star,\rm MT}}{M_{\odot}}\right)^{-1}\\ 
    &\ \ \ \ \ f_{\rm c}e_{\rm MT}^2(1-e_{\rm MT})^{23/2}.
    \end{aligned}
    \label{eq:MdotTidest0}
\end{equation}

\noindent 

\begin{figure*}
    \centering
    \includegraphics[width=\textwidth]{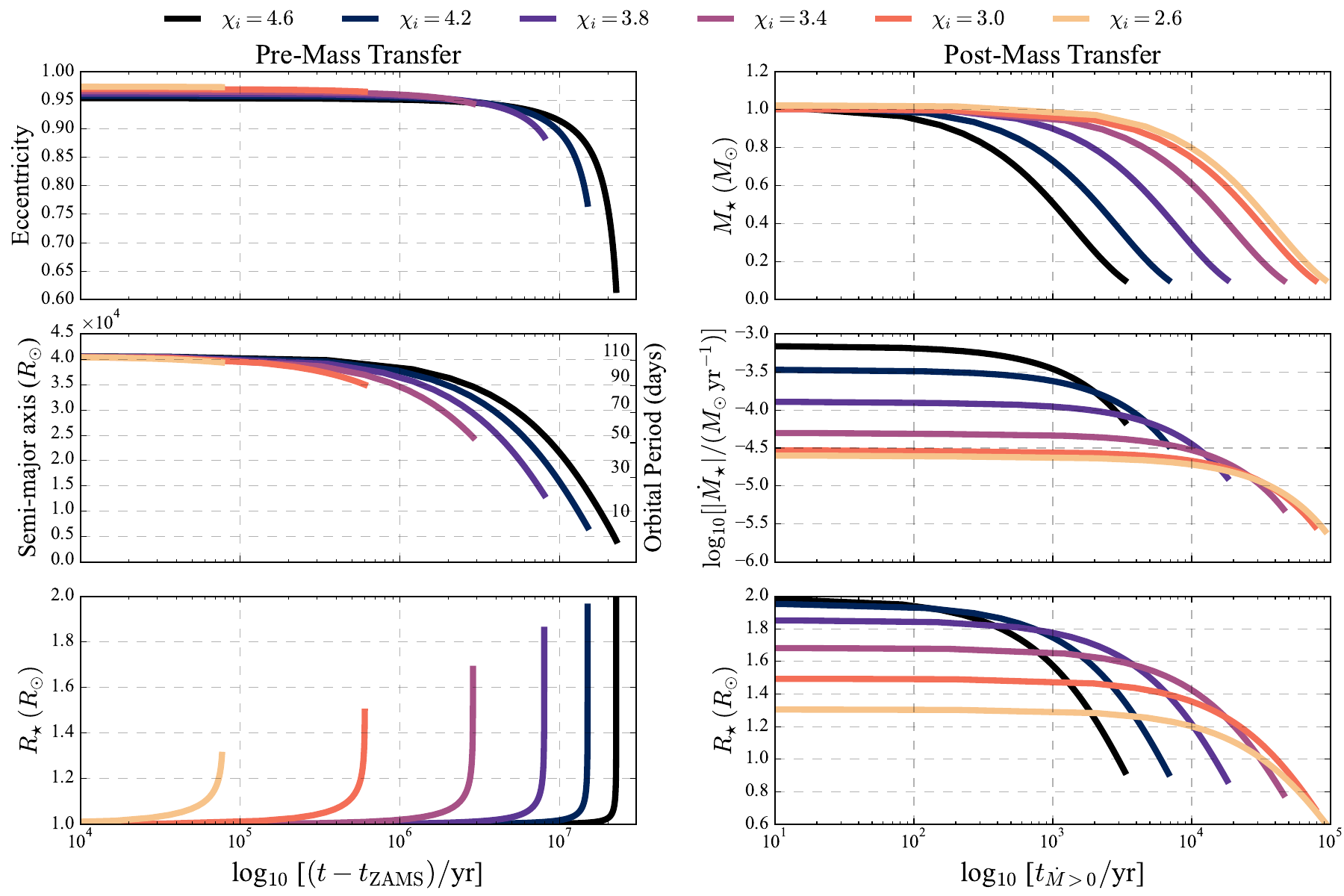}
    \caption{\textit{Stars on Highly Eccentric Orbits:} Numerical solution for the evolution of orbital and stellar properties (eqs. \ref{eq:num_GW1} \& \ref{eq:num_GW2}) for stars orbiting a $7\times 10^7 M_{\odot}$ SMBH before (\textit{left}) and after (\textit{right}) the onset of mass transfer at $r_{\rm p} = a(1-e) = 2r_{\rm t}$. The models shown are for a 1$R_{\odot}$ \& $1M_{\odot}$ star at $t_{\rm ZAMS}$, have an initial orbital period $P_i = 114.2$days, and a range of initial tidal heating parameters $\chi_i \equiv \rp/\rtidal$, which is equivalent to a range of initial eccentricities.  Prior to mass transfer the orbit decays due to GWs and the stellar radius increases due to tidal heating.   The mass transfer rate is set by the tidal heating timescale, and is significantly higher than predicted by gravitational wave orbital decay.}
    \label{fig:num_para}
\end{figure*}

\begin{figure*}
    \centering
    \includegraphics[width=\textwidth]{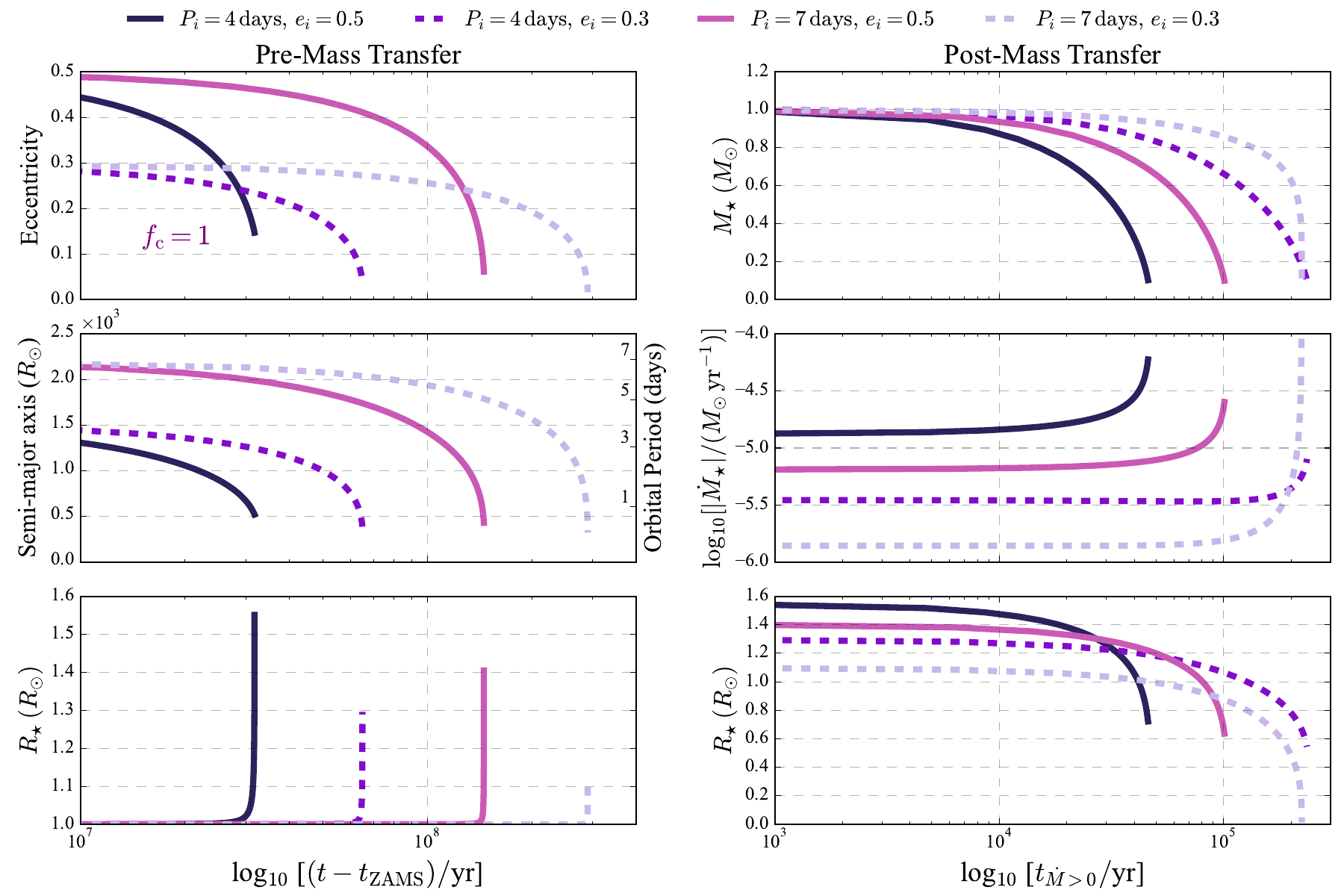}
    \caption{\textit{Stars on Near-circular Orbits ($f_{\rm c} = 1$):} Numerical solution for the evolution of orbital and stellar properties (eqs. \ref{eq:num_GW1} \& \ref{eq:num_GW2}) for stars orbiting a $3\times 10^6 M_{\odot}$ SMBH before (\textit{left}) and after (\textit{right}) the onset of mass transfer at $r_{\rm p} = a(1-e) = 2r_{\rm t}$. Here $f_c$ is the dimensionless normalization of the tidal heating rate in eq. \ref{eq:Edot_circ}; this figure shows our fiducial choice $f_c=1$. The models shown are for a 1$R_{\odot}$ \& $1M_{\odot}$ star at $t_{\rm ZAMS}$ and have an initial eccentricity $e_i = 0.3$ (dashed lines) or $e_i = 0.5$ (solid lines) and an initial orbital period $P_i$ of 4 or 7 days. As in Figure \ref{fig:num_para}, the mass transfer rate is again largely set by the tidal heating timescale, but for the most circular orbits this is only a few times higher than $\dot M_{\rm GW,circ}$ derived in eq. \ref{eq:MdotGW}.}
    \label{fig:num_circ}
\end{figure*}

\begin{figure*}
    \centering
    \includegraphics[width=\textwidth]{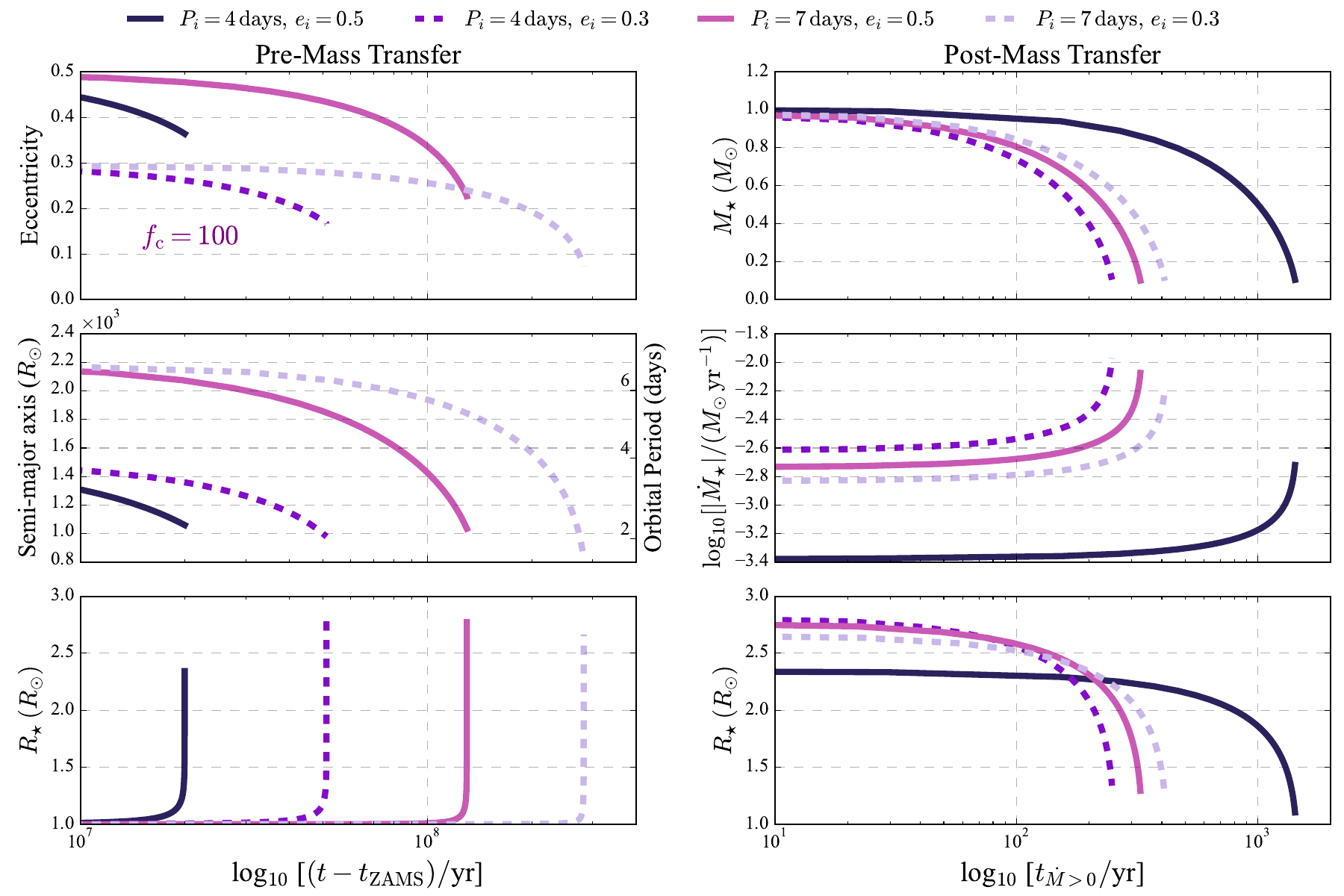}
    \caption{\textit{Stars on Near-circular Orbits ($f_{\rm c} = 100$):} Numerical solution for the evolution of orbital and stellar properties (eqs. \ref{eq:num_GW1} \& \ref{eq:num_GW2}) for stars orbiting a $3\times 10^6 M_{\odot}$ SMBH before (\textit{left}) and after (\textit{right}) the onset of mass transfer at $r_{\rm p} = a(1-e) = 2r_{\rm t}$. The tidal heating is our fiducial model (eq. \ref{eq:Edot_circ}) multiplied by $f_c = 100$ to explore the impact of larger tidal heating rates. The models shown are for a 1$R_{\odot}$ \& $1M_{\odot}$ star at $t_{\rm ZAMS}$ with an initial eccentricity $e_i = 0.3$ (dashed lines) or $e_i = 0.5$ (solid lines) and an initial orbital period $P_i$ of 4 or 7 days. For the larger tidal heating rate considered here, the star inflates much more quickly, leading to mass transfer at a higher eccentricity and a significantly higher resulting mass transfer rate $\sim 10^{-3} {\rm M_\odot \, yr^{-1}}$.  Such a model could be the progenitor of short-period quasi-periodic eruptions such as eRO-QPE2.}
    \label{fig:num_circ_fc100}
\end{figure*}

When scaled to properties of typical solar-mass stars, the stable mass transfer rate estimated including tidal heating (Eq. \ref{eq:MdotTides}) is notably larger than simple estimates based on gravitational wave orbital decay (Eq. \ref{eq:MdotGW}).  This strongly suggests that the properties of stable mass transfer for stars orbiting massive black holes are set by mass loss initiated by tidal inflation of the stellar radius rather than by gravitational wave-induced orbital decay.   This conclusion is subject to the non-negligible uncertainty in the tidal heating rate, but the large difference in the numerical factor in front of eq. \ref{eq:MdotTidest0} relative to eq. \ref{eq:MdotGW} suggests that it is robust (recall again that $f_p \gtrsim 1$ and $f_c \gtrsim 1$ are quite plausible).   A corollary of this conclusion is that the stable mass transfer phase is likely to last at most of order $10^3-10^4\rm \, yrs$ (and possibly less) instead of $10^6-10^7\rm \, yrs$ based on gravitational wave orbital decay. 

Tidal heating transfers energy from the orbit to the star, which can in turn modify the orbital properties.
For highly eccentric orbits, the dominant effect is the change in orbital period, whereas for near-circular orbits it is the change in eccentricity. Their timescales at the onset of mass transfer are
\begin{equation}
\begin{aligned}
\left(\frac{P}{\dot P}\right)_{\rm tides,para} &\simeq \frac{10^6}{f_{\rm p}}{\rm yrs}\frac{R_{\star,\rm MT}}{R_{\odot}}\left(\frac{M_{\star,\rm MT}}{M_{\odot}}\right)^{-1}\\ 
&\ \ \ \left(\frac{\MBH}{10^6M_{\odot}}\right)^{2/3}\left(\frac{P_{\rm MT}}{100\rm days}\right)^{1/3}\\
t_{\rm tides,circ} &\sim \frac{e}{\dot e} \simeq \frac{1.9\times 10^5}{f_{\rm c}}\, {\rm yrs} \left(\frac{M_{\star,\rm MT}}{M_{\odot}}\right)^{4/3} \\ 
&\ \ \ \ \left(\frac{\MBH}{10^6M_{\odot}}\right)^{2/3}\left(\frac{R_{\star,\rm MT}}{R_{\odot}}\right)^{-1}.
\label{eq:ttide}
\end{aligned}
\end{equation}
Equation \ref{eq:ttide} shows that orbital evolution by tides is 2-3 orders of magnitude slower than the corresponding mass transfer timescales ($\sim M_\star/\dot M_{\star}$) evaluated from Equation \ref{eq:MdotTidest0}. Thus, once mass transfer commences, orbital evolution by both tides and gravitational waves is generally slower than the evolution of the star's mass. For our fiducial parameters ($f_{\rm p} = f_{\rm c} = 1$), gravitational wave orbital decay dominates over tides for the orbital evolution prior to the onset of mass transfer, but tidally-induced orbital changes will likely be more important if $f_{\rm p}\gg1$ or $f_{\rm c}\gg1$.

\subsection{Mass, Radius, and Orbital Evolution with \ \ Gravitational Waves and Tidal Heating} \label{sec:tides_GW}

Here we combine tidal heating and gravitational wave orbital decay for a solar-mass main sequence star to develop a simple model for the orbital and stellar evolution from the onset of stellar expansion due to tidal heating, to the onset of mass transfer from the star to the BH, to the end of the stable mass transfer phase.   The calculations here {\em assume} stable mass transfer and do not follow the internal stellar structure, and so cannot directly assess the validity of this assumption.  We return to this point in \S \ref{sec:MESA}. Orbital evolution here is given by gravitational waves \citep{Peters1964} as well as tides (although tides are subdominant for our fiducial parameters per eq. \ref{eq:ttide}).   For the latter we consider tidal circularization for nearly-circular orbits but include both $\dot E_{\rm tides}$ and $\dot J_{\rm tides}$ for the highly eccentric case, with $\dot J_{\rm tides} \simeq \dot E_{\rm tides}/2 \Omega_p$.

As in \S \ref{sec:tides_noGW}, we assume mass transfer is initiated and maintained at $\rp \simeq 2\rtidal$, such that
\begin{equation}
R_{\star} = \frac{a(1-e)}{2\MBH^{1/3}}M_{\star}^{1/3};
\label{eq:RM}
\end{equation}
in this case, combining with Equation \ref{eq:tideode2}, we can solve for the evolution of the stellar radius using
\begin{equation}
\frac{d{\rm ln}R_{\star}(t)}{dt} = \ \ \frac{\dot E_{\rm tides}}{E_{\star}\phi_{\rm E}}\\
\label{eq:num_GW1}
\end{equation}
when $\dot{M} = 0$, and
\begin{equation}
\frac{d{\rm ln}R_{\star}(t)}{dt} = \frac{6}{5}\frac{\dot{a}}{a}-\frac{6}{5}\frac{\dot{e}}{1-e}-\frac{\dot E_{\rm tides}}{5E_{\star}\phi_{\rm E}},
\label{eq:num_GW2}
\end{equation}
when $\dot{M} > 0$. Here, $\dot E_{\rm tides}$, $\dot a$, and $\dot e$ are functions of the evolving orbital and stellar parameters.

Figures \ref{fig:num_para} \& \ref{fig:num_circ} show the solutions to Equations \ref{eq:num_GW1} \& \ref{eq:num_GW2} before (\textit{left}) and after (\textit{right}) the onset of mass transfer at $\rp = 2\rtidal$ for highly eccentric and near-circular orbits, respectively. Runaway expansion of the stellar radius before mass transfer is evident in both cases since the tidal heating rate is a strong function of the stellar radius. For both cases, we note that the tidal orbital evolution timescale ($\tau_{\rm tides}$) always remains subdominant when compared to $\tau_{\rm GW}$, which is itself small compared to $\tau_{\rm heat}$ when mass transfer begins. Hence, orbital evolution by tides is not very important for these fiducial parameters.

\subsubsection{Highly Eccentric Orbits}

For the solar-type stars on highly eccentric orbits in Figure \ref{fig:num_para}, we assume an initial orbital period of $114.2$ days around a $7\times10^7 M_{\odot}$ SMBH, inspired by estimated parameters of ASASSN-14ko \citep{Payne2021}. The qualitative evolutionary behavior shown does not depend strongly on the exact fiducial choice of parameters. We initialize the orbit starting at a range of tidal heating parameters ($\chi_i$), which also corresponds to a range of initial eccentricities ($e\sim 0.955-0.975$).

As expected, stars with smaller $\chi_i$, i.e., smaller initial pericenter radii, quickly inflate due to tidal heating and rapidly reach $\chi = 2$, when mass transfer begins. For stars with larger initial pericenter distances (larger $\chi_i$), the tidal heating is initially smaller; the longer initial tidal heating timescale means that there is more significant orbital decay of the semi-major axis due to gravitational wave emission prior to the onset of mass transfer. In all cases, after $\sim 10^4-10^7$ yrs, the stable mass transfer phase commences.  The timescale for the onset of mass-transfer in the left panels of Figure \ref{fig:num_para} is largely set by the time it takes tidal energy deposition to increase the stellar radius by order unity given the initial value of $\chi_i$ (we showed in Eqs. \ref{eq:tgrav} \& \ref{eq:ttide} that gravitational wave and tidal changes to the orbit operate on yet longer timescales). All the models shown here thus implicitly require the star to be placed on a long-period orbit with $\chi_i \sim 2-5$ by a mechanism such as the Hills mechanism \citep[e.g.][]{Hills1988, Cufari2022, Lu2023, linialSari2023}.   

As we derived in equations \ref{eq:MdotTides} \& \ref{eq:MdotTidest0}, the mass transfer rate $\dot M_{\star}$ decays exponentially and depends only on the initial orbital period at mass transfer. Also consistent with the analytic estimates presented in \S \ref{sec:tides_noGW}, the tidal heating timescale $\tau_{\rm heat} \sim E_{\star}/\dot E_{\rm tides}$ remains much shorter than the gravitational wave timescale and determines the stable mass transfer rate.

\subsubsection{Near-circular Orbits}

For solar-type stars on mildly eccentric orbits in Figure \ref{fig:num_circ}, we adopt a lower mass for the SMBH ($\MBH = 3\times10^6 M_{\odot}$), inspired by the estimated parameters of QPE hosts \citep{Wevers2022}. We test four cases, each representing a combination of initial eccentricity (0.3 or 0.5) and orbital period (4 or 7 days), which allow all the systems to evolve to $\rp = 2\rtidal$ before their main-sequence lifetime concludes. All models are subject to tidal heating following Equation \ref{eq:Edot_circ}, start with a high $\chi_{\rm i}$ (and hence low $\dot E_{\rm heat,c}$), and undergo significant orbital decay to milder eccentricities due to gravitational wave circularization. The final semi-major axes at mass transfer ($a_{\rm MT}\sim 330-520 R_\odot$) correspond to orbital periods of 0.4 - 0.8 days.

Similarly to the high-eccentricity case, stars on orbits with the shortest $a_{\rm i}$ inflate the fastest due to tidal heating and reach $\chi =2$. Models with a longer $a_{\rm i}$ require a longer heating time, undergo more appreciable orbital decay, and end up transferring mass at lower eccentricities. This is because, unlike the highly eccentric calculations in Figure \ref{fig:num_para}, gravitational waves circularize the mildly eccentric orbits much more rapidly within  $\sim 10^7-10^8$yrs, until tidal heating takes over to dominate the subsequent evolution. The mass transfer rate $\dot M_{\star} \sim 10^{-6}-10^{-4} M_{\odot}/\rm yr$ is set again by the tidal heating timescale,
although for most of the evolution it is only a factor of $\sim 10-100$ times larger than the mass-transfer rate from GWs alone (this is because $e^2$ is small at the onset of mass transfer, making eq. \ref{eq:MdotTidest0} more similar to eq. \ref{eq:MdotGW}). We terminate the calculation when $M_{\star}$ falls below $0.1M_{\odot}$, where many models have already started to exhibit a diverging mass transfer rate expected from eq. \ref{eq:MdotTides}. 

We also show a case with a significantly higher tidal heating rate ($f_{\rm c} = 100$) in Figure \ref{fig:num_circ_fc100} with the same set of initial parameters as in Figure \ref{fig:num_circ}. As argued in the discussion below eq. \ref{eq:Edot_circ}, this level of heating is plausible. Here, the runaway expansion of the stellar radii is even more rapid, which leads to an earlier onset of mass transfer, less significant orbital evolution, and greater $R_{\star, \rm MT}$ compared to the $f_{\rm c} = 1$ case in Figure \ref{fig:num_circ}. The final semi-major axes at mass transfer ($a_{\rm MT}\sim 830-1050 R_\odot$) correspond to orbital periods of 1.6 - 2.3 days. Despite a significantly shorter tidal circularization timescale, the orbital evolution is still largely dominated by gravitational-wave orbit decay.  By the onset of mass transfer, however, tidal heating again dominates the evolution, setting the much higher stable mass transfer rate of $\dot M_{\star} \sim 10^{-4}-10^{-2} M_{\odot}/\rm yr$ (as expected given $\dot M_\star \propto f_c$ in eq. \ref{eq:MdotTidest0}).

\begin{figure*}
    \centering
    \includegraphics[width=\textwidth]{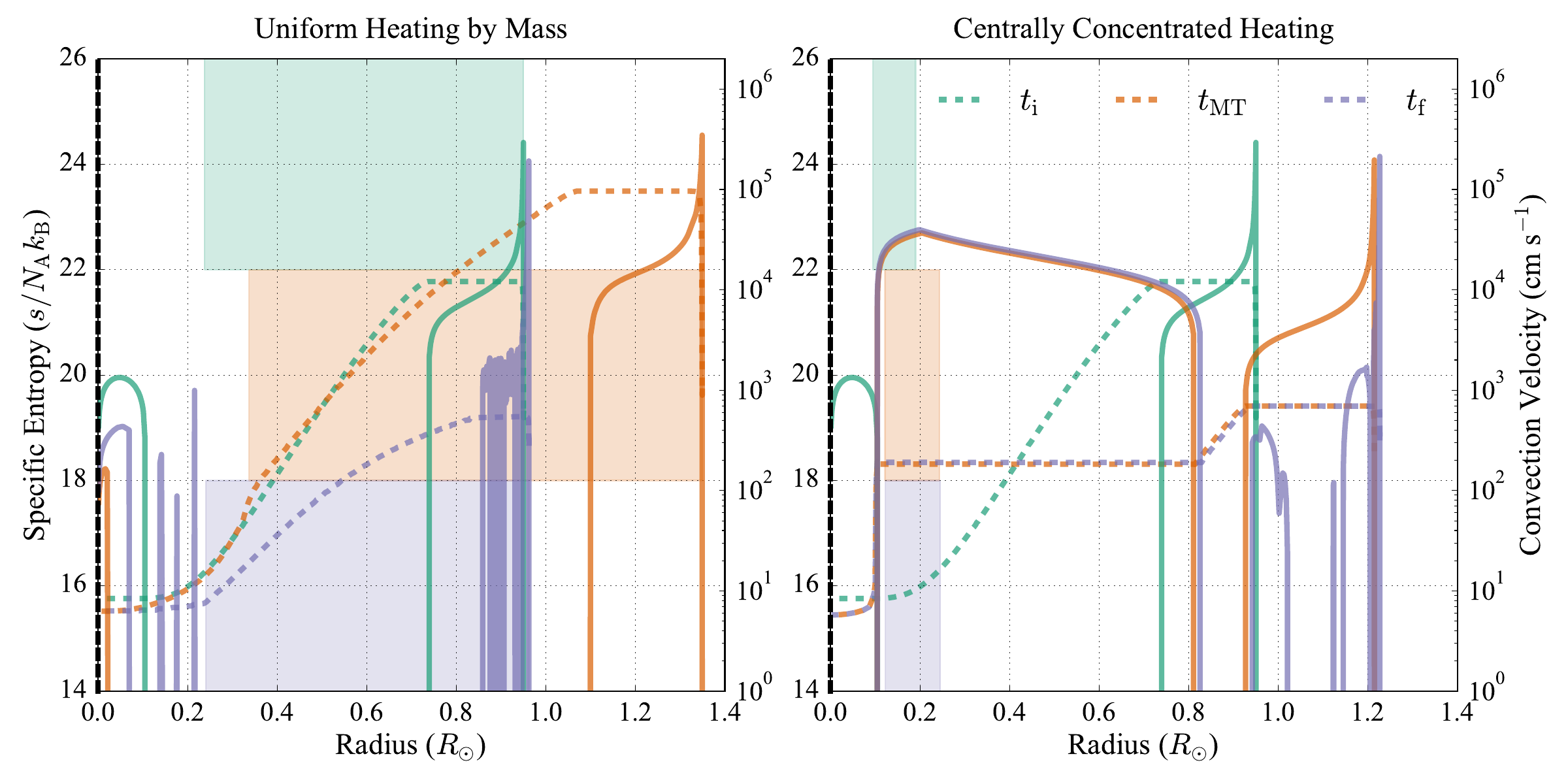} 
    \caption{Internal stellar profiles of tidally heated stars (shown here: mildly eccentric, $e_i = 0.5$, $P_i = 7$ days, violet line in Fig. \ref{fig:MESA_QPE_fixP}) with the dashed lines showing specific entropy and solid lines showing convective velocity within the star. The \textit{left} and \textit{right} panels correspond to the (1) centrally concentrated and (2) uniformly throughout the star cases discussed in \S \ref{sec:MESA}. The 3 different colors shown represent the stellar structure at different stages of stellar evolution: the green, orange, and purple profiles correspond to models closest to zero-age-main-sequence ($t_{\rm i}$), beginning of mass transfer ($t_{\rm MT}$), and after unstable mass transfer sets in ($t_{\rm f}$). The shaded regions represent the radial extent within which tidal heating energy is being deposited at each stage. {Uniform heating throughout the star inhibits convection and allows for a phase of stable mass transfer, whereas centrally concentrated heating generates large convective regions, leading to unstable mass transfer.}}
    \label{fig:stellar_profile}
\end{figure*}

\begin{figure*}
    \centering
    \includegraphics[width=\textwidth]{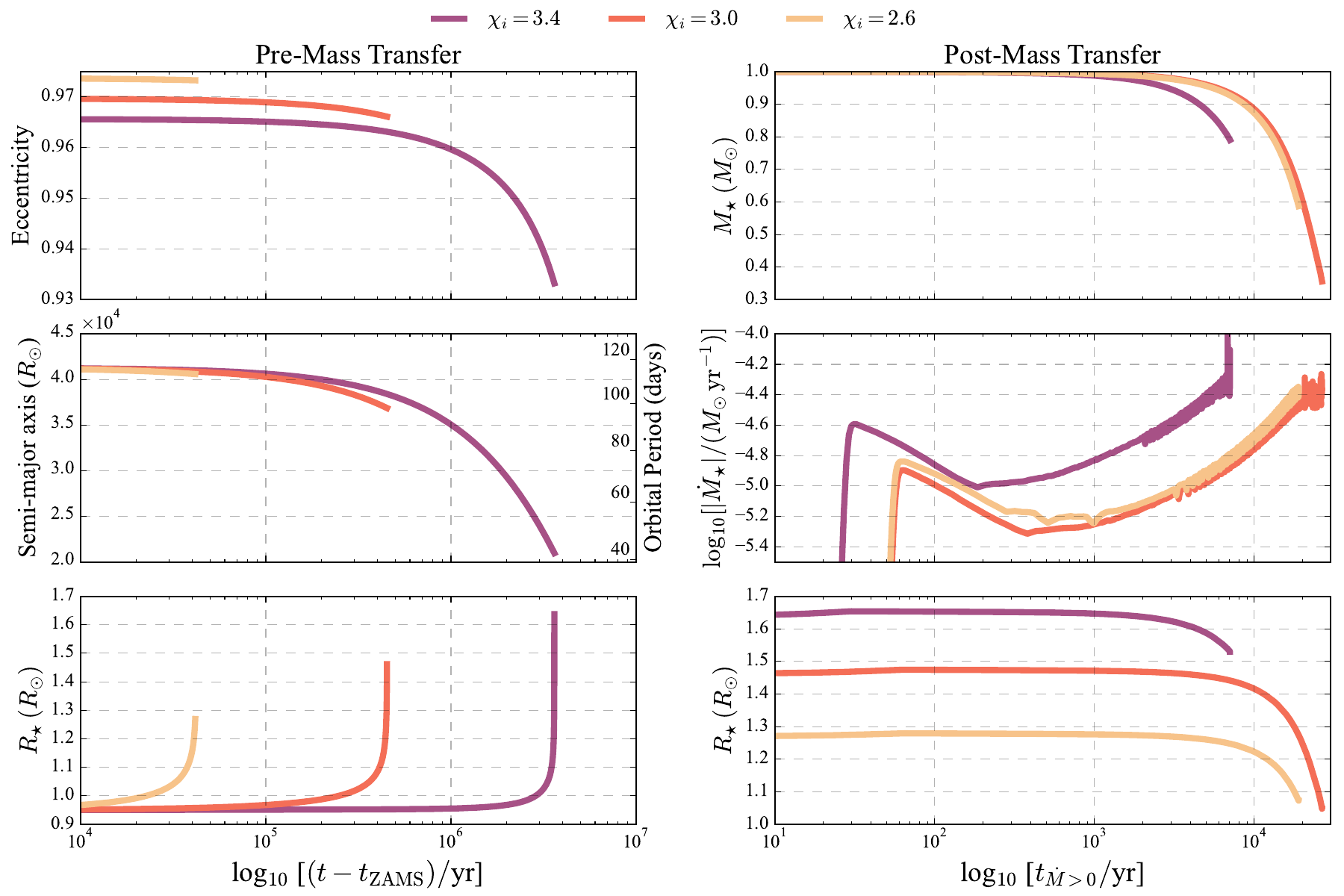}    
    \caption{\textit{Stars on Highly Eccentric Orbits:} MESA solution for the evolution of orbital and stellar properties (eqs. \ref{eq:num_GW1} \& \ref{eq:num_GW2}) before (\textit{left}) and after (\textit{right}) the onset of mass transfer at $r_{\rm p} = a(1-e) = 2r_{\rm t}$. The models are subsets of those for which Figure \ref{fig:num_para} shows our simplified analytic model: a $1M_{\odot}$ star at ZAMS orbiting a $7\times 10^7 M_{\odot}$ SMBH, an initial orbital period $P_i = 114.2$days, and a range of initial tidal heating parameters $\chi_i \equiv \rp/\rtidal$, which is equivalent to a range of initial eccentricities. We generally recover similar orbital and stellar evolution to models evolved analytically in Fig. \ref{fig:num_para}.  The MESA models end with the onset of unstable mass transfer, as indicated by the rapid rise in $\dot M_{\star} $ in the right panel at the latest times.}
    \label{fig:MESA_pTDE}
\end{figure*}

\section{Numerical Models with MESA}
\label{sec:MESA}

The analytic model presented in \S \ref{sec:analytic} provides simple scaling relationships for the stable mass transfer rate as a function of stellar and orbital parameters, and the tidal heating model.  However, it does not capture the stability of mass transfer or how tidal heating changes the internal structure of the star. Here, we perform more detailed calculations of tidally heated stars orbiting SMBHs with the Module for Experiments in Stellar Astrophysics \citep[MESA,][version r24.08.1]{Paxton2011, Paxton2013, Paxton2015, Paxton2018, Paxton2019, Jermyn2023}. All our binary MESA models consist of a point mass SMBH orbited by a 1 $M_{\odot}$ zero-age main sequence (ZAMS) star\footnote{The inlists can be found at: \url{https://zenodo.org/records/19440898}.}. 

Following the arguments summarized in the introduction, in our MESA simulations we assume that the angular momentum and energy lost from the star during mass-transfer are not returned to the orbit.
By default, MESA does not handle gravitational wave orbital decay for eccentric orbits (circular options are included in MESA). Hence, we add $dL/dt$ and $de/dt$ terms\footnote{In MESA, this corresponds to \texttt{jdot\_extra\_routine} and \texttt{edot\_extra\_routine} in \texttt{run\_binary\_extras}} following \citet{Peters1964} in all our orbit evolution with additional terms described in \S \ref{sec:tides_GW} to include effects of tidal orbital evolution (the latter are negligible for the calculations presented here). Furthermore, for tidal heating, we deposit energy in the star based on its evolving stellar and orbital properties. The spatial distribution of tidal heating in stellar interiors is uncertain and depends on the detailed mechanism of tidal dissipation. Such dissipation is plausibly non-linear for the large tidal distortions considered here. We will consider two possibilities in our MESA models that bracket the stellar response to tidal heating in a manner described below, both with uniform heating per unit mass\footnote{We avoid heating  the stellar core at the smallest radii to prevent creating an artificial steep temperature gradient, which we found can lead to small numerical timesteps.}:
\begin{enumerate}
    \item Centrally Concentrated: $0.1 < r/R_{\star} < 0.2$. This is motivated by non-linear wave breaking of internal gravity waves due to geometric focusing in stellar cores \citep{Goodman1998}. Centrally concentrated heating also drives convection, promoting unstable mass transfer. 
    \item Uniform (per unit mass) in the outer 75\% of the star: $0.25 < r/R_{\star}$. While this is less directly physically motivated, applying constant heating per unit mass raises the entropy more in the cooler outer regions, which inhibits convection and thereby stabilizes mass transfer.
\end{enumerate}

As in Section \ref{sec:analytic}, we use Equations \ref{eq:Edot_para} \& \ref{eq:Edot_circ} to calculate the heating power associated with stars in highly eccentric orbits and mildly eccentric orbits, respectively. We note that this time-averaged approximation is more accurate for persistent heating in the latter case, while highly eccentric orbits should receive heating predominantly near each pericenter passage in short bursts with the large-amplitude tidally excited waves damped by non-linear processes \citep{Kumar1996}.

For our MESA models, we consider the mass transfer stable or unstable when the stellar radius decreases or increases in response to decreasing mass, respectively. For unstable mass transfer, the increase in radius drives a sharp increase in $\dot M_{\star}$, further increasing the stellar radius. This produces a runaway process which MESA is unable to solve numerically, leading to a very small timestep. The onset of unstable mass transfer is difficult to explicitly see in the radius evolution panels in Figures \ref{fig:MESA_pTDE} \& \ref{fig:MESA_QPE_fixP}. However, a sharp rise in $\dot M_{\star}$ is evident at the end stage of each stellar model, and is a signature of the onset of unstable mass transfer.

Figure \ref{fig:stellar_profile} shows the convective velocity and specific entropy profiles for our two chosen heating models for one of our MESA calculations with the star on a mildly eccentric orbit ($e_i = 0.5, \, P_i = 7$ days). Three snapshots are shown: near  ZAMS ($t_{\rm i}$), at the onset of mass transfer ($t_{\rm MT}$), and the final model near the onset of unstable mass transfer ($t_{\rm f}$). All of our centrally heated models quickly become fully convective when the tidal heating rate becomes comparable to the stellar luminosity (Fig. \ref{fig:stellar_profile}). Once $\chi \simeq 2$, mass transfer is immediately unstable: the star expands in response to mass loss, and MESA cannot continue to find solutions.  On the contrary, when heating is evenly distributed per unit mass in most of the star, the star remains mostly radiative with a low-mass convective envelope on the exterior. Since the centrally heated models undergo unstable mass transfer, we focus our analysis on the more uniform heating models in what follows. 

\subsection{Highly Eccentric Orbits} 

We adopt the same configuration as in our analytic model in \S \ref{sec:analytic} with a $\MBH = 7\times 10^7 M_{\odot}$ point mass SMBH orbited by Sun-like ZAMS stars with an initial orbital period of 114.2 days. Figure \ref{fig:MESA_pTDE} shows the stellar and orbital evolution for a range of initial tidal heating parameters $\chi_i = 2.6 - 3.4$, which also translates to a range of initial eccentricities ($e_i \sim 0.965 - 0.975$). For our MESA models, heating is gradually turned on for numerical stability, and reaches the full value in Equation \ref{eq:Edot_para} starting 100 years after $t_{\rm ZAMS}$. 

The \textit{left} column in Figure \ref{fig:MESA_pTDE} shows the stellar radius and orbital evolution before the onset of mass transfer, measured as a function of time since ZAMS. The \textit{right} column shows the evolution of stellar mass, mass transfer rate, and stellar radius after the star begins to lose mass. For this range of $\chi_i$, MESA can always find a stable mass transfer solution that lasts for $\sim 10^4$yrs. Subsequently, the stellar evolution culminates when the radius increases as mass loss continues, driving a sharp increase in $\dot M_{\star}$, at which point MESA can no longer continue to model the dynamics of this runaway process.

Overall, we obtain stable mass transfer rates from MESA models similar to our analytics in Section \ref{sec:analytic},  set by the tidal heating timescale instead of the gravitational wave timescale. Compared to our analytic model, which assumes stable mass transfer at all times, the star in our MESA model loses $\sim 20$\%-70\% of its mass before the mass transfer becomes unstable (in the case of uniform heating per unit mass throughout most of the star). Models with higher $\chi_i$ tend to have higher $\dot M_{\star}$, and are more prone to unstable mass transfer for our phenomenological tidal heating model.

\subsection{Mildly Eccentric Orbits}

\begin{figure*}
    \centering
    \includegraphics[width=\textwidth]{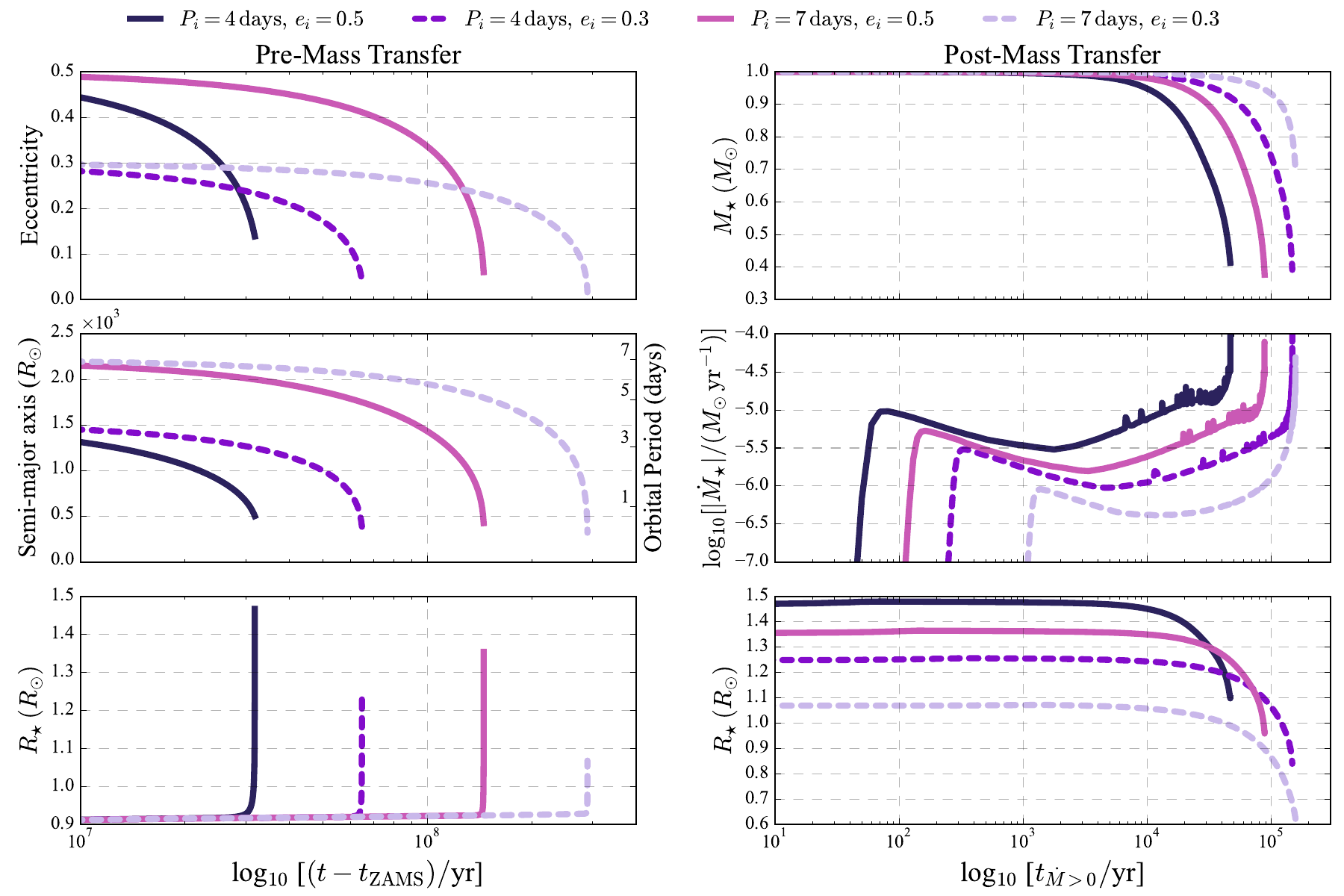}
    \caption{\textit{Stars on Near-circular Orbits:} MESA solution for the evolution of orbital and stellar properties before (\textit{left}) and after (\textit{right}) the onset of mass transfer at $r_{\rm p} = a(1-e) = 2r_{\rm t}$. The models shown are the same as for the analytic model in Figure \ref{fig:num_circ}: a $1M_{\odot}$ star at ZAMS orbiting a $3\times 10^6 M_{\odot}$ SMBH, an initial eccentricity $e_i = 0.3$ (dashed lines) or $e_i = 0.5$ (solid lines) and an initial orbital period $P_i$ of 4 or 7 days. We recover similar orbital and stellar evolution to models evolved analytically in Fig. \ref{fig:num_circ}. The MESA models end with the onset of unstable mass transfer, as indicated by the rapid rise in $\dot M_\star$ in the right panel at the latest times.}
    \label{fig:MESA_QPE_fixP}
\end{figure*}

For stars on mildly eccentric orbits, as in \S \ref{sec:analytic}, we adopt a $\MBH = 3\times 10^6 M_{\odot}$ point mass SMBH orbited by Sun-like ZAMS stars with a range of eccentricities and orbital periods. Figure \ref{fig:MESA_QPE_fixP} shows stellar and orbital evolution with the same initial orbital parameters as in Figure \ref{fig:num_circ}: initial eccentricity 0.3 or 0.5 and orbital period of 4 or 7 days. Likewise, the \textit{left} column in Figure \ref{fig:MESA_QPE_fixP} shows the stellar radius and orbital evolution before the onset of mass transfer, measured as a function of time since ZAMS, and the \textit{right} column shows the evolution of stellar mass, mass transfer rate, and stellar radius after the star begins to lose mass. 

As we found in our analytic model (\S \ref{sec:analytic}), the mass transfer rate here is set by the tidal heating timescale $\dot M_{\star} \sim 10^{-6} - 10^{-4} M_{\odot}/\rm yr$, which results in a $\sim 10^5$-year phase of stable mass transfer during which a Sun-like star can lose 40-60\% of its mass before unstable mass transfer sets in. The behavior of $\dot M_{\star}$ evolution in the MESA runs is also similar to our analytic calculations, where $\dot M_{\star}$ always evolves secularly higher, until the mass transfer eventually becomes unstable, driving a sharp increase in $\dot{M}$. 

\vspace{1cm}

\section{Discussion \& Summary}
\label{sec:summary}

Stars in galactic nuclei can experience significant tidal heating due to their proximity to the central SMBH. For stars whose orbits reach sufficiently small pericenter distances relative to the tidal radius, the heating eventually causes runaway expansion of the star, and the star is destined to transfer mass to the SMBH. Motivated by the interplay of these physical processes, we developed an analytic model that describes the stable mass transfer solution in the presence of tidal heating, in which a star's intrinsic response to mass loss is to shrink sufficiently in radius (unstable mass transfer is also possible, as we discuss below).  The specific tidal heating models we use are based on solar mass stars, but many qualitative features of our evolutionary models are more broadly applicable. We then incorporated orbital changes due to gravitational waves and tides into a semi-analytic model of coupled stellar and orbital evolution.  In parallel, we performed the same calculations using more realistic stellar models in MESA, with models for tidal heating included in the MESA calculations.  The MESA models explicitly address whether mass transfer is stable or unstable since the internal stellar structure determines whether the star expands or contracts in response to mass loss.  Our primary results, their implications, and avenues for future work are discussed below.

In standard models of stable mass transfer in binary systems, the mass transfer rate is set by the rate of angular momentum loss in the system, which drives the binary together (e.g., by gravitational wave emission or magnetic braking of stellar rotation).   We have shown that for stars orbiting SMBHs there is a new stable mass transfer solution set by the tidal heating timescale $\tau_{\rm heat} \sim E_{\star}/\dot E_{\rm tides}$, in which $\dot M_{\star} \sim M_\star/\tau_{\rm heat}$.  We analytically derived the properties of this stable mass transfer solution (eqs. \ref{eq:MdotTides}, \ref{eq:MdotTidesConstants} \& \ref{eq:MdotTidest0}; Fig. \ref{fig:num_para}, \ref{fig:num_circ} \& \ref{fig:num_circ_fc100}) and verified it in MESA models (Fig. \ref{fig:MESA_pTDE} \& \ref{fig:MESA_QPE_fixP}).   Our results show that the tidal heating timescale is shorter than the gravitational wave inspiral time unless the eccentricity at the onset of mass transfer is $\ll \sim 0.01$.  Thus, stable mass transfer with tidal heating has a larger mass-transfer rate than mass transfer induced by gravitational wave orbital decay, significantly reducing the lifetime of the mass-transferring phase and increasing the luminosity of star-powered accretion in galactic nuclei. 

 For stars on highly eccentric orbits, our analytic model predicts that, if mass transfer is stable, it occurs for $10^3-10^5$ years with $\dot M_{\star,\rm para}(t_{\rm MT}) \sim 10^{-5}-10^{-3} M_{\odot}/\rm yr$; the mass transfer rate scales inversely with the orbital period at the onset of mass transfer and linearly with the tidal heating rate (eq. \ref{eq:MdotTides}).   Such systems would likely appear observationally as repeating nuclear transients in galactic nuclei, with emission powered by accretion of stellar debris lost at pericenter and/or circularization shocks associated with such debris.    Our default models predict mass-transfer rates a factor of $\sim 10$ lower than needed to explain the luminosities of a system like ASASSN-14ko \citep{Payne2021} which has been exhibiting persistent flares every $\simeq 114$ days over the past decade (ASASSN-14ko requires $\langle \dot M \rangle \sim 3\times 10^{-3} M_{\odot} \rm yr^{-1}$ to power the flares by accretion). This suggests that many RNTs powered by tidally heated mass loss may be fainter than the pTDE candidates detected thus far. Different stellar models can, however, have significantly different tidal heating rates (e.g., \citealt{Lee1986} for polytropes), so a larger heating rate for some stellar models and/or evolutionary phases is plausible. The eccentric stable mass transfer model is essentially a gentler version of the partial-TDE picture, which has previously been used to explain the multi-wavelength flares of ASASSN-14ko \citep[e.g.][]{Payne2021, Cufari2022, Bandopadhyay2024}.  The two models differ in the following ways:   tidal heating-induced mass transfer applies to stars whose pericenter distance approaches the tidal radius slowly as in the empty loss cone regime, or for stars that are dynamically placed on bound orbits with $2 \lesssim r_p/r_t \lesssim 5$.   By contrast, if stars are scattered onto orbits with $1 \lesssim r_p/r_t \lesssim 2$, the calculations here are not applicable and instead the star undergoes a true partial tidal disruption.  In the latter regime, the tidal heating generated mass loss is likely small compared to the matter directly stripped by the tidal forces \citep{Bandopadhyay2025}. Tidal heating-induced mass transfer is more difficult to simulate, but would be very useful to study directly with multi-dimensional simulations.   The simulations to date that come the closest are those of \citet{Bandopadhyay2024} and \citet{Liu2025} who studied repeated interactions between stars and a SMBH accounting for the tidally induced change to the stellar structure.  The regime explored in this paper differs in that the initial stellar pericenter distance is larger, the star is tidally inflated prior to the onset of mass-transfer, the mass-loss per pericenter passage is generally lower, and the star will be roughly co-rotating with the orbit at pericenter by the onset of mass-transfer (though we do not explicitly model rotation). 
 
For stars on near-circular orbits, our analytic model with a fiducial tidal heating prescription predicts that, if mass transfer is stable, it occurs for $\sim 10^4 - 10^6$ years with $\dot M_{\star,\rm circ} \sim 10^{-6}-10^{-4} M_{\odot}/\rm yr$; the mass transfer rate scales with orbital eccentricity squared and linearly with the tidal heating rate (eq. \ref{eq:MdotTidest0}).  For sufficiently small eccentricity at the onset of mass transfer, the tidal heating timescale is longer than gravitational wave orbital decay, and thus the latter sets the stable mass transfer rate.   Observationally, such stable mass transfer on a roughly circular orbit could power a persistent low-luminosity AGN.   Indeed if the rate of producing stars on mildly eccentric orbits around SMBHs is $\sim 10^{-6}-10^{-5} \,{\rm yr^{-1}}$ for a Milky Way mass galaxy \citep{linialSari2023, Lu2023}, a significant fraction of galaxies could host accretion powered by a mass-transferring star.

Interactions between the star and the disk it produced can also potentially power episodic flares once or twice per orbit \citep{Lu2023}. In particular, since the star's orbit is likely inclined to the equatorial plane of the SMBH spin, the disk formed by stellar mass transfer becomes aligned with the SMBH spin by the Bardeen-Petterson effect \citep{Bardeen1975}. The star and its tidally stripped stellar debris will then collide with the disk it created on an inclined orbit, producing phenomenology similar to that found in recent work on QPEs  \citep[e.g.][]{LinialMetzger2023,Tagawa2023,Yao2024}.  Our default tidal heating models predict stable mass transfer rates lower than what is necessary to explain the quiescent (between flare) emission in QPE sources such as GSN 069 and eRO-QPE2.  
As we have stressed throughout the paper, however, the tidal heating rate that sets the stable mass transfer rate in our models is uncertain.  For example, inertial wave excitation can produce tidal heating rates that are several orders of magnitude larger than the internal gravity wave excitation implemented in this work \citep[e.g.][]{Barker2020,LQ2024}. It is also plausible that non-linear dissipation of the equilibrium tide energy (e.g., \citealt{Weinberg2012}) may become important at the large tidal energies that necessarily accompany mass transfer around SMBHs.   If there are channels for stable mass transfer via tidal heating that produce $\dot M_{\star,\rm circ} \sim 10^{-3}-10^{-2} M_{\odot}/\rm yr$ (which requires tidal heating rates $\sim 100$ times larger than our default model; see eq. \ref{eq:MdotTidest0} and Fig. \ref{fig:num_circ_fc100}), such a channel could account for the shortest period QPEs such as GSN 069 and eRO-QPE2 \citep{Miniutti2019, Arcodia2021}.\footnote{Previous models of QPEs invoked unstable mass transfer to explain the high accretion rates needed in QPE models \citep{Lu2023,linialSari2023}.  By showing that the stable mass transfer rate can be much larger in the presence of tidal heating, our results open up the space of progenitor models for QPEs to include stable mass transfer.  This does not, of course, preclude that unstable mass transfer channels could still be viable as well.} 

Recent discoveries of longer-period QPEs are associated with optically selected TDEs \citep{Nicholl2024}.   By contrast,  some of the shortest period QPEs, such as GSN 069 and eRO-QPE2, have quiescent (non-flaring) emission that is quite long-lived \citep{Miniutti2023, Miniutti2023b} and it is less clear if these are standard tidal disruption events (TDEs; see, however, \citealt{Guolo2025} who argue that GSN 069 is plausibly a TDE with an unusually long disk viscous time).   The stable mass transfer model qualitatively explains this long-duration disk emission as well as the lack of other standard AGN features such as broad lines.   It can also naturally explain the rapid onset of the quiescent emission because the accretion rate increases to the stable mass transfer value on the time it takes tidal heating to inflate the star a distance comparable to the scale-height of the stellar photosphere $H$,  $\tau_{\rm MT} \sim \tau_{\rm heat} (H/R_{\star})$, where $\tau_{\rm heat}$ is the tidal heating timescale.  If $\tau_{\rm heat} \sim 100-1000$ years, as needed to explain the quiescent disk emission in QPEs, $\tau_{\rm MT} \sim 0.1-1$ years is sufficiently short to explain the rapid onset of accretion in these systems.  A quiescent disk fed by mass-loss from a star has an unusual optical-UV spectrum distinct from that of TDEs that may allow direct tests of this model \citep{Lu2023} (for GSN 069, \citealt{Guolo2025b} favor a standard spreading accretion disk UV spectrum from $\simeq 5-10$ eV but the difference between the star-fed disk and a spreading disk at those wavelengths is modest; see Fig. 4 of \citealt{Lu2023}; it is more pronounced at longer wavelengths). If mass-transferring systems indeed power some QPEs, it can only produce shorter period systems ($\lesssim$ days).  A prediction is that such shorter recurrence time systems would have systematically different quiescent disk emission than QPEs following optically selected TDEs, and should lack the early time optical emission seen in TDEs. 

There is still large uncertainty as to where tidal heating energy is deposited inside a star for the large tidal amplitudes of interest in this work.  Our MESA models are consistent with expectations that if tidal heating generates large convective regions, it will lead to unstable mass transfer. This is realized in our model by tidal heating concentrated in the center of a Sun-like star.  By contrast, constant heating per unit mass throughout most of the star increases the size of the radiative zone (Fig. \ref{fig:stellar_profile}). This leads to an extended phase of stable mass transfer in MESA models. The true outcome of mass transfer in tidally heated stars will require more detailed calculations that properly evolve the tidal heating rate and its spatial distribution in the stellar interior as a function of time in response to the changing stellar structure. We defer this to future work. 

The work presented here should be improved in a number of ways.   Indeed, we hope to have been forthright that our primary goal has been to assess the qualitatively new features that result from tidal heating in stars orbiting close to SMBHs.    Many of the quantitative details of our predictions are uncertain and will require significantly more work.   This includes the tidal heating rates themselves as well as the spatial distribution of the tidal heating in the stellar interior.   In addition,   our models of highly eccentric stable mass transfer are particularly idealized.    Our MESA implementation approximates the tidal energy deposition as uniform throughout the entire orbital period, while in reality the tidal energy is deposited near pericenter and then thermalized by non-linear interactions.   The mass loss itself will be impulsive as well, concentrated near pericenter.   Whether such mass transfer is indeed stable for some stellar progenitors remains to be seen, but we suspect that our results about the stability of mass transfer for a given stellar structure are robust (the biggest uncertainty is then the structure of tidally heated stars). For example, the calculations of \citet{Bandopadhyay2025} of stellar response to impulsive mass stripping in the context of repeating transients are consistent with the established expectation for stability also found in this paper:  \citet{Bandopadhyay2025}  find that lower mass stars, which have adiabatic indices comparable to structural polytropic indices, expand with mass loss while higher mass [radiative] stars contract. This is equivalent to our result that fully convective stars (with similar adiabatic and structural polytropic indices) undergo unstable mass transfer while radiative stars undergo stable mass transfer. 

A further limitation of our calculations is that they are hydrostatic and thus do not capture possible time-dependent pulsations or dynamical mass ejection once the stellar response becomes strongly non-linear. This caveat may be particularly relevant for donors with longer dynamical times, such as red giants, or during the unstable mass-transfer phase. In related 1D hydrodynamic \texttt{MESA} calculations of heated red-giant envelopes in the common-envelope context, \citet{clayton2017} found that sustained energy injection can drive large-amplitude pulsations and episodic shell ejections. In our problem, however, the hydrodynamic response is likely sensitive to the spatial distribution of the tidal heating within the star, which is itself highly uncertain. A meaningful exploration of such effects therefore requires a more self-consistent treatment of both the magnitude and location of tidal energy deposition, and we leave this to future work.

Finally, our calculations neglect the interaction between the orbiting star and accretion disks formed in (unrelated) parabolic TDEs, many of which occur on the timescales of interest in this paper \citep{LinialMetzger2023}.   It is unclear if the star approaching mass transfer can in fact survive its repeated collisions with TDE disks \citep{LinialMetzger2023,Linial2024b,Yao2024}.

\section*{Conclusions}

We have studied the evolution of tidally heated Sun-like stars orbiting SMBHs using analytic estimates and 1D \texttt{MESA} calculations.  As in the related problem of binary star systems, the resulting mass transfer can be either stable or unstable depending on the properties of the tidal heating and the stellar structure.   Our main conclusions are as follows:
\begin{itemize}
    \item Tidal heating can inflate stars well before they would begin mass transfer in a GW-driven picture alone. As a result, when stable mass transfer occurs, its rate is typically set by the tidal heating timescale rather than the GW orbital-decay timescale.

    \item For the systems considered here, the resulting stable mass-transfer rates are plausibly $\sim 10^{-3}-10^{-5}\,M_\odot\,{\rm yr}^{-1}$ (and in some cases larger), sufficient to power low-luminosity AGN. The corresponding stable phase is short, typically $\sim 10^{3}-10^{4}$ yr rather than the much longer lifetime expected for purely GW-driven evolution.

    \item The stability of mass transfer depends sensitively on where tidal energy is deposited within the star. In our \texttt{MESA} models, centrally concentrated heating tends to drive the star toward convection and unstable mass transfer, whereas more spatially extended heating can keep the star radiative and lead to a stable mass-transfer phase.

    \item Tidally heated mass transfer may explain some repeating nuclear transients, including possible progenitors of partial TDE candidates and short-period QPEs. However, our results suggest that many stars undergoing tidal-heating induced mass transfer are likely fainter than the currently known population and remain to be discovered.

    \item A key uncertainty in our models is the tidal heating physics itself, including both the total heating rate and where the dissipated energy is deposited inside the star.   Unlike in binary star systems in our problem the tidal heating can change the structure of the star, modifying the stability of mass transfer and the tidal heating itself.     
    A more self-consistent treatment of the coupled problem of tides, stellar structure, and orbital evolution will be needed to determine more robustly when mass transfer is stable or unstable.
\end{itemize}

\begin{acknowledgements}
We thank Eric Coughlin, Jim Fuller, Itai Linial, Wenbin Lu, Brian Metzger, Selma de Mink, Aleksandra Olejak, and Enrico Ramirez-Ruiz for useful conversations, and Mathieu Renzo and Sunny Wong for helpful MESA suggestions.  We appreciate the referee's thorough report that improved the paper significantly.   This work was supported in part by a Simons Investigator grant from the Simons Foundation (EQ).   This research benefited from interactions at workshops funded by the Gordon and Betty Moore Foundation through grant GBMF5076 and through interactions at the Kavli Institute for Theoretical Physics, supported by NSF PHY-2309135.
\end{acknowledgements}

\bibliographystyle{mn2e}
\bibliography{main}

\end{document}